\documentclass[referee]{mn2e}
\newcommand{\be}{\begin{eqnarray}}
\newcommand{\ee}{\end{eqnarray}}

\newcommand{\Ms}{M_{\star}}

\newcommand{\Mp}{M_p}
\newcommand{\Md}{M_d}
\newcommand{\MJ}{M_{\rm J}}

\usepackage{graphicx}
\usepackage{amsmath}
\graphicspath{{figures/}}

\begin{document}

\title[The effects of disc warping on the inclination of planetary
orbits]{The effects of disc warping on the inclination of planetary
  orbits}

\author[Caroline Terquem] { Caroline Terquem \\ Department of Astrophysics,
  University of Oxford, Keble Road, Oxford OX1 3RH, UK \\
  Institut d'Astrophysique de Paris, UPMC Univ Paris 06, CNRS,
  UMR7095, 98 bis bd Arago, F-75014, Paris, France \\ E-mail:
  caroline.terquem@astro.ox.ac.uk }

\date{}

\pagerange{\pageref{firstpage}--\pageref{lastpage}} \pubyear{}

\maketitle

\label{firstpage}

%
%

\begin{abstract}

  The interaction between a planet located in the inner region of a
  disc and the warped outer region is studied.  We consider the stage
  of evolution after the planet has cleared--out a gap, so that the
  planetary orbit evolves only under the gravitational potential from
  the disc.  We develop a secular analysis and compute the evolution
  of the orbital elements by solving Lagrange's equations valid to
  second order in the eccentricity.  We also perform numerical
  simulations with the full disc potential.  In general, the
  interaction between the disc and the planet leads to the precession
  of the orbit.  The orbital plane therefore becomes tilted relative
  to the disc's inner parts, with no change in the eccentricity.  When
  the inclination approaches 90 degrees, there is an instability and
  the eccentricity increases.  In this case, both the inclination and
  the eccentricity develop large variations, with the orbit becoming
  retrograde.  As the eccentricity reaches high values, we would
  expect tidal capture on a short orbit of the planet by the star to
  occur.  This instability happens when the disc is severely warped,
  or if there is a significant amount of mass in a ring inclined by at
  least 45 degrees relative to the initial orbital plane.  The
  inclination of the orbit does not depend on the semimajor axis nor
  on the planet's mass.  However, for a significant inclination to be
  generated on a timescale of at most a few Myr, the planet should be
  beyond the snow line.  The process described here would therefore
  produce two distinct populations of inclined planets: one with
  objects beyond the snow line with at most moderate eccentricities,
  and another with objects on short circularized orbits.

\end{abstract}

\begin{keywords}
celestial mechanics --- planetary systems --- planetary systems:
formation --- planetary systems: protoplanetary discs --- planets and
satellites: general
\end{keywords}

%
%

\section{Introduction}
\label{sec:intro}

As of today, about 250 extrasolar planets have been detected through
both radial velocity and transit measurements.  The combination of
these two detection methods makes it possible to determine the angle
between the sky projection of the stellar spin axis and the orbit
normal (that we will hereafter call the {\em projected inclination
  angle}) using the Rossiter--McLaughlin effect.  Among the 70 planets
for which this angle has been obtained, about 30 show significant
misalignment, with a few even displaying retrograde orbits.  The
distance of most of these planets to their host star is below 0.1~au,
and most of these objects have a mass larger than half that of
Jupiter.  The sample is therefore highly biased towards massive and
close planets.  However, these detections indicate that misalignement
is not rare.

If planets on short orbits have formed in and migrated through a disc,
they would be in a plane misaligned with that of the star's equator
only if the disc itself were misaligned, or if interactions had taken
place that would put the planets on misaligned orbits.  Such
interactions, occuring after dissipation of the disc, include secular
perturbation by a distant stellar or planetary companion that produces
Kozai cycles with tidal friction (Fabrycky \& Tremaine 2007; Wu,
Murray \& Ramsahai 2007; Naoz et al. 2011), planet--planet scattering
(Chatterjee et al.  2008) or secular chaos (Wu \& Lithwick 2011).
Processes leading to the misalignment of the disc itself include
accretion  (at later times) of material with angular momentum
misaligned with that of the star (Bate, Lodato \& Pringle 2010), close
encounter with a star while the disc accretes from an extended
envelope (Thies et al.  2011), or precession induced by the
interaction with a stellar companion in a binary system (Batygin
2012).  Planetary orbits inclined with respect to the stellar
equatorial plane would also result if the stellar spin axis were
tilted due to interaction with the disc (Foucart \& Lai 2011; Lai,
Foucart \& Lin 2011).

The interactions studied so far, which would misalign an orbit
initially in the stellar equatorial plane, are presumed to take place
after the disc has dissipated.  Note, however, that if the orbital
plane were inclined relative to the disc while it is still present,
the orbit would not necessarily (re)align as a result of the
frictional force exerted by the disc on the planet when it crosses it.
This is because the secular gravitational interaction between an
inclined planet and the disc's outer parts leads to Kozai cycles that
maintain either the inclination or the eccentricity at large values,
thus delaying alignment with the disc and circularization of the orbit
(Terquem \& Ajmia 2010; Teyssandier, Terquem \& Papaloizou 2013).
This Kozai effect could not be seen in previous simulations of planets
on inclined orbits in which the disc was truncated at rather small
radii (e.g., Marzari \& Nelson 2009), but was observed in the SPH
simulations of Xiang-Gruess \& Papaloizou (2013) which used a more
extended disc and higher orbital inclinations.

So far, the mechanisms that have been put forward to produce a
misalignment of a planetary orbit relative to a disc rely on the
interaction of the planet with at least one stellar or planetary
companion.  Here, we investigate the interaction of the planet with
the disc itself, and show that a warping of the disc's outer parts may
lead to a significant inclination of the planet's orbit relative to
the disc's inner parts.

\subsection{Warping of protoplanetary discs}

The outer parts of protostellar discs may become warped as a result of
tidal interaction with a binary star (Terquem \& Bertout 1993, 1996;
Papaloizou \& Terquem 1995).  Most stars are in multiple systems
(e.g., K\"ohler 2008) and the distribution of the separation of
pre--main sequence binaries has a peak around 30 au (Leinert et al.
1993).  It is therefore expected that tidal interactions between a
circumstellar disc and the companion are common.  If the disc and the
orbit are not coplanar, such interactions lead to a warping of the
disc's outer parts.  Evidence for warped discs have been seen, e.g.,
in the young triple system WL~20 (Barsony 2002) and in the Bok globule
CB~26 (Launhardt \& Sargent 2001).  Also, in the young binary system
HK~Tau, the disc around one of the stars is observed not to be
coplanar with the orbital plane (Stapelfeldt et al. 1998), which
implies that it may be warped.  

Warping may also occur if the disc accrete at later times material
with angular momentum misaligned with that of the star (Bate, Lodato
\& Pringle 2010).  This is expected as stars form within clusters in
which gravitational interactions are important.  It was suggested by
Tremaine (1991) that the obliquities of the outer planets in our solar
system could have been produced by a change in the direction of the
angular momentum of the planetary system after it had formed.  The
proposed causes for this change were a time-dependant angular momentum
of the material falling into the disc, possibly due to inhomogeneities
in the collapsing core, or gravitational torques due to mass
concentrations within the core.  Whether asymmetric infall might
actually result in a warped disc or not is still a matter of debate.
The accretion of material with misaligned angular momentum onto the
disc outer parts does produce an inclination of these outer parts 
with respect to the inner regions, but such an inclination may be
smoothed out by viscous diffusion through the disc.  Whether the final
configuration is planar or not depends on whether the viscous
diffusion timescale is smaller or larger than the precession timescale
(see discussion below).  In the simulations by Bate, Lodato \& Pringle
(2010), the disc would settle into a planar configuration for the
range of parameters used, but whether it is always the case or not is
still an open question.

It has been proposed that, in a stellar binary system in which a disc
is misaligned with respect to the orbital plane, planets forming into
this disc would themselves be inclined relative to the central star's
equatorial plane (Batygin 2012).  This is because the disc is tilted
due to the precession of the disc's angular momentum vector around the
orbital angular momentum vector.  However, warping was neglected in
this study, under the argument that a self--gravitating disc similar
to the one considered would only precess as a rigid body if it
remained unwarped (see also Batygin et al. 2011).

There have been extensive studies of the dynamics of both galatic and
protostellar discs that have shown that, for a disc to precess as a
rigid body in a non spherically symmetric potential, it actually has
to bend to alter the precession frequency at each radius so that the
rate is everywhere the same (Sparke \& Casertano 1988; Hofner \&
Sparke 1994).  The disc therefore settles into a discrete bending mode
(representing a warp) that is referred to as the modified tilt mode
because, in the limit that the external potential is spherically
symmetric, it reduces to the trivial rigid--tilt mode.  Differential
precession can be smoothed out by gravitational torques in a
self--gravitating disc (Toomre 1983), and also by radial pressure
forces and viscous diffusion (Papaloizou \& Terquem 1995, Larwood et
al. 1996).  These processes are efficient if the timescale on which
the different parts of the disc communicate with each other is shorter
than the precession timescale.  In a protostellar disc in which
pressure forces are likely to dominate over self-gravitating forces,
rigid body precession is achieved if the sound crossing time through
the disc is much smaller than the precession period, so that bending
waves propagate through the disc sufficiently fast for the different
parts of the disc to adjust their precession rate to a constant value.
Communication becomes diffusive in a disc in which turbulent stresses
dominate.

In general, in a disc dominated by pressure forces, rigid body
precession of the whole disc is achieved with only a small degree of
warping (Larwood et al. 1996, Larwood 1997).  In that case, as
envisionned by Batygin (2012), the relative inclination of the disc
with the central star's equatorial plane is due mainly to the
precession, not to the warp.  However, it was found in the numerical
simulations of Larwood et al. (1996) that the disc could maintain
itself as a long--lived and coherent structure even when a strong warp
develops and the inner and outer parts of the disc precess at
different rates.  In that case, there is a coupling between the two
parts of the disc due to pressure gradients, and the rigidly
precessing outer parts drag the inner parts behind it.

Planets forming in the disc's inner parts would therefore interact
with the strongly misaligned outer parts, and this is the type of
situation we investigate in this paper.  

\subsection{Setup}

We consider a planet that has formed in the inner parts of a disc
which outer parts have been warped according, possibly, to one of the
processes described in the previous section.  


Note that, if the disc's warp were due to a companion on a noncoplanar
orbit in a binary system, gravitational interaction between this
companion and the planet would affect the orbital evolution of the
planet.  This is not taken into account in this paper, where we focuss
on the interaction between the warped disc and the planet
independently of the cause of the warp.  The dynamics of a planet in a
noncoplanar binary system, taking into account disc warping and the
presence of the companion, will be the subject of a further paper.

We consider the stage of evolution after an inner cavity has been
opened in the disc, and we assume that the planet's orbit lies in this
cavity.  Inner holes are commonly observed in the so--called
transitional discs (Andrews et al.  2013).  They are usually assumed
to be opened either by X--ray photoevaporation (Owen et al. 2011) or
by tidal interaction with low mass object(s).  A giant planet is
expected to clear--out a gap, so that the parts of the disc interior
to the orbit are no longer replenished efficiently from the outer disc
and accrete onto the central star on a viscous timescale if the disc
is active.  Note that, in this scenario, the planet may be expected to
migrate in as a result of tidal interaction with the outer parts of
the disc.  However, migration would be slow if the mass of gas in the
vicinity of the planet were significantly smaller than the mass of the
planet itself.  Both planet's migration and accretion of the disc
material would also be very inefficient if the disc were passive in
the region where the planet is located.  In that case, the planet
would move in a gap but there would still be material interior to its
orbit.

In the calculations presented in this paper, we assume for simplicity
that there is no mass interior to the planet's orbit, but our results
would hardly be changed if only a gap were cleared, with still
material orbiting in the region inside the orbit.  This is because the
gravitational interaction with the disc is dominated by the parts of
the disc which are outside the orbit (Terquem \& Ajmia 2010).  The
important point here is that we assume there is no friction between
the disc and the planet, so that the planetary orbit evolves only
under the gravitational potential from the disc.

\subsection{Plan of the paper}

In section~\ref{sec:interaction}, we develop a secular analysis of the
interaction between a planet and a warped disc.  We first write the
gravitational potential exterted by the disc on the planet and
describe the disc model used in this study in
section~\ref{sec:gravitation}.  In section~\ref{sec:expansion}, we
expand the potential in the case where the radius of the inner edge of
the disc is large compared to the planet's semimajor axis, and
calculate the secular terms as a function of the orbital elements to
fourth order in the eccentricity in section~\ref{sec:secular}.
Lagrange's planetary equations, valid to second order in the
eccentricity, are then written in section~\ref{sec:Lagrange}.  We
solve these equations in section~\ref{sec:solutions}, both for a flat
and a warped disc, to obtain the evolution of the orbital elements.
In the case of a flat disc, we recover the results of Terquem \& Ajmia
(2010), who showed that perturbation by an outer flat disc inclined
relative to the planetary orbit may lead to Kozai cycles.  In the case
of a warped disc, we find that, in general, the interaction bewteen
the planet and the disc leads to the precession of the planetary
orbit.  The orbit therefore becomes inclined relative to the disc's
inner parts, with no increase of its eccentricity.  We find that there
is an instability when the inclination approaches $90^{\circ}$.  In
that case, the eccentricity starts to increase and both the
inclination and the eccentricity have very large amplitude variations,
with the orbit becoming retrograde.
In section~\ref{sec:simulations}, we perform simulations that enable
us to follow the evolution of the orbit when the eccentricity gets
large.  Finally, in section~\ref{sec:discussion}, we summarize and
discuss our results.

%
%

\section{Secular analysis}
\label{sec:interaction}


\subsection{Gravitational potential and disc model}
\label{sec:gravitation}

We consider a planet of mass $\Mp$ orbiting around a star of mass
$\Ms$ which is itself surrounded by a disc of mass $\Md$.  We denote
by $(x,y,z)$ a Cartesian coordinate system centred on
the star and $(r, \theta, z)$ the associated cylindrical polar
coordinates.  
The warped disc has its central parts tangent to the $(x,y)$--plane
and its elevation above this plane is taken to be:
\be
z(r, \theta) = \left( \frac{r}{R_o} \right)^p H_o \sin \theta,
\label{zwarp}
\ee

\noindent where $R_o$ is the disc's outer radius, $H_o$ is the
maximum elevation at $r=R_o$, and $p>1$ is an integer which controls
the warp's curvature ($p=1$ would correspond to a disc rigidly tilted
relative to the $(x,y)$--plane).  We introduce $\alpha$, the inclination of
the disc's outer edge with respect to the $(x,y)$--plane, so that: \be
H_o= R_o \tan \alpha,
\label{Ho}
\ee

\noindent and let $R_i$ be the disc's inner edge.  Initially, the orbit of
the planet is in the $(x,y)$--plane and has a semimajor axis $a<R_i$.
We suppose that the angular momentum of the disc is large compared to
that of the planet's orbit so that the effect of the planet on the
disc is negligible.

Note that if the disc as a whole is tilted with respect to the
stellar equatorial plane, the $(x,y)$--plane does not coincide
with the equatorial plane of the star.  This would be the case if what
caused the warp also induced a precession of the disc (e.g.,
fly--by of a star).

The gravitational potential exerted by the disc at the location
of the planet is:

\be \Phi_d = -G \int_{R_i}^{R_o} \Sigma(r) r dr \int_{0}^{2\pi}
\frac{d\theta}{\sqrt{r^2+r_p^2 - 2rr_p \cos \left( \theta - \theta_p
    \right) + \left[ z_p - \left( \frac{r}{R_o} \right)^p R_o \tan \alpha
    \sin \theta \right]^2}},
\label{potential}  \ee

\noindent where the subscript $p$ refers to the planet, $G$ is the
gravitational constant and $\Sigma$
is the mass density in the disc.  We assume:

\be
\Sigma(r)= \Sigma_0 \left( \frac{r}{R_o} \right)^{-q},
\label{sigma}
\ee

\noindent where $q$ is a positive number.  In the numerical
simulations described below, we will set $q=1/2$.  The coefficient
$\Sigma_0$ can be expressed in term of the disc mass $M_d$:

\be \Sigma_0 = \frac{(-q+2) M_d}{ 2 \pi R_o^2 \left[ 1 - \left(R_i/R_o
    \right)^{-q+2} \right]}.  \ee

The total potential at the location of the planet has four
contributions: one from the disc, which is given by
equation~(\ref{potential}), one from the central star, an indirect
contribution since the center of the coodinate system
is accelerated, and another indirect contribution since
that, in general, the $(x,y)$--plane rotates as a result of the disc's
precession.  However, these last three terms do not yield any secular
contribution and will be ignored in this section.

\subsection{Expansion of the gravitational potential}
\label{sec:expansion}

We define $u \equiv r_p/r$, $v \equiv z_p/r$ and $m \equiv \left(
  r/R_o \right)^{p-1} \tan \alpha$.  The potential in
equation~(\ref{potential}) can then be written in the form:

\be
\Phi_d = -G \int_{R_i}^{R_o} \Sigma(r) dr {\cal I} \left( r,u,v \right) ,
\label{potential2}
\ee

\noindent with:

\be {\cal I} \left(r,u,v \right) = \int_0^{2 \pi} \frac{ d
  \theta}{\sqrt{1+u^2-2u \cos \left( \theta - \theta_p \right) + \left( v -
    m \sin \theta \right)^2}} . 
\label{Idefinition}
\ee

We assume that $R_i \gg r_p$ and $R_i \gg z_p$ so that $u$ and $v$ are
small compared to unity everywhere in the disc.  To second--order in
$u$ and $v$, ${\cal I} \left( r,u,v \right)$ can then be expanded as:

\be {\cal I} \left( r,u,v \right) = {\cal I} \left( r,0,0 \right) +
\frac{u^2}{2} \frac{\partial^2 {\cal I}}{\partial u^2} \left( r,0,0 \right)
 + \frac{v^2}{2} \frac{\partial^2 {\cal I}}{\partial v^2} \left( r,0,0
\right)  + u v \frac{\partial^2 {\cal I}}{\partial u \partial
  v} \left( r,0,0 \right) ,
\label{Iexpansion}
\ee

\noindent where the omitted first--order terms vanish.  It is shown in the
appendix~\ref{appendixA} that this expansion yields the following
expression for the potential~(\ref{potential2}):

\be \Phi_d = -\frac{G \Sigma_0}{R_o} \left( {\cal I}_1 R_o^2 + {\cal I}_2
    r_p^2  + {\cal I}_3 r_p^2 \cos 2 \theta_p 
   + {\cal I}_4 z_p^2  + {\cal I}_5 r_p z_p \sin
    \theta_p  \right) ,
\label{potential3}
\ee

\noindent where ${\cal I}_1, {\cal I}_2, {\cal I}_3, {\cal I}_4$ and
${\cal I}_5$ are dimensionless integrals over $r$ which depend only on
disc parameters.  Their expression in term of elliptic
integrals is given in appendix~\ref{appendixA}.

\subsection{Secular terms}
\label{sec:secular}

We wish to calculate the long--term effect of the interaction between
the disc and the planet, obtained by averaging the
potential~(\ref{potential3}) over the mean longitude of the planet. In
order to perform this averaging, we first express the planet's polar
coordinates $r_p, \theta_p$ and $z_p$ in terms of the planet's orbital
elements.  We denote by $a$, $e$, $I$, $\omega$ and $\Omega$ the
semimajor axis, eccentricity, inclination, argument of pericentre and
longitude of ascending node, respectively, of the orbit, defined with
respect to the reference frame $(x,y,z)$.  We then have (Murray \&
Dermott 1999):

\begin{eqnarray}
r_p \cos \theta_p & = & R \left[ \cos \Omega \cos \left( \omega + f \right)
- \sin \Omega \sin \left( \omega + f \right) \cos I \right] , 
\label{rpcos} \\
r_p \sin \theta_p & = & R \left[ \sin \Omega \cos \left( \omega + f \right)
+ \cos \Omega \sin \left( \omega + f \right) \cos I \right] , 
\label{rpsin}
\\
z_p & = & R \sin \left( \omega + f \right) \sin I,
\label{zp}
\end{eqnarray} 

\noindent where $R$ is the distance between the planet and the star
center and $f$ is the true anomaly of the planet (polar angle in the
orbital plane relative to the pericentre).  To perform a time
averaging of the potential~(\ref{potential3}), we further express $R$,
$\sin f$ and $\cos f$ in terms of the mean anomaly $M$, which is $2
\pi$--periodic and a linear function of time.  Expansions of $R$, $\sin
f$ and $\cos f$ to fourth order in $e$ are given in
appendix~\ref{appendixB}.  We insert these series expansion in
equations~(\ref{rpcos}), (\ref{rpsin}) and~(\ref{zp}) above, and 
expand $r_p^2$, $r_p^2 \cos 2 \theta_p $, $z_p^2$, $r_p
z_p \sin \theta_p$, and therefore the potential~(\ref{potential3}), in
terms of $M$ to fourth order in $e$.  Finally we perform a time
average to obtain the secular potential $\left< \Phi_d \right>$.
Time--averaging eliminates from the expansion the terms which are odd
in $e$.  As it happens, the coefficients of $e^4$ also vanish.  The
series expansions we obtain are therefore valid up to fifth order in $e$.
The time--averages of $r_p^2$, $r_p^2 \cos 2 \theta_p $,
$z_p^2$, $r_p z_p \sin \theta_p$, which are needed to compute $\left<
  \Phi_d \right>$ from equation~(\ref{potential3}), are given below:

\begin{eqnarray}
  \frac{ \left< r_p^2 \right>}{a^2} & = & \frac{1}{4} \left( 3 + \cos 2I
  \right) + \frac{e^2}{8} \left[ 3 \left( 3 + \cos 2I \right) + 10
    \cos 2 \omega \sin^2 I \right] + {\cal O}(e^6),
\label{rp2}
\end{eqnarray}

\begin{eqnarray}
  \frac{ \left< r_p^2 \cos 2 \theta_p \right>}{a^2} = & & 
  \frac{1}{2} \cos 2 \Omega \sin^2 I \nonumber \\ & + &
  \frac{e^2}{8} \left[ 5 \left( 3 + \cos 2I \right) 
    \cos 2 \omega \cos 2 \Omega + 6
    \cos 2 \Omega \sin^2 I - 20 \cos I \sin 2 \omega \sin 2 \Omega \right] 
\nonumber \\
  & + & {\cal O}(e^6),
\label{rp2cos}
\end{eqnarray}

\begin{eqnarray}
\frac{ \left< z_p^2 \right>}{a^2} & = & \frac{1}{2} \sin^2 I +
\frac{e^2}{4} \left( 3 - 5 \cos 2 \omega \right) \sin^2 I +  {\cal O}(e^6),
\label{zp2}
\end{eqnarray}

\begin{eqnarray}
  \frac{ \left< r_p z_p \sin \theta_p \right>}{a^2} = & & \frac{1}{4} 
\sin 2I \cos \Omega  \nonumber \\
& + & \frac{e^2}{4} \sin I \left( 3 \cos I \cos \Omega - 
5 \cos I \cos 2 \omega \cos \Omega + 5 \sin 2 \omega \sin \Omega \right) 
\nonumber \\
& + & {\cal O}(e^6).
\label{rpzpsin}
\end{eqnarray}

\subsection{Lagrange's planetary equations}
\label{sec:Lagrange}

In secular theory, the semimajor axis $a$ is constant, and the
time variation of $e$, $\Omega$, $\omega$ and $I$ are given by the
Lagrange's planetary equations (e.g., Roy 1978):

\begin{eqnarray}
\frac{de}{dt} & = & - \frac{ \sqrt{1-e^2}}{n a^2 e}
\frac{\partial {\cal R}}{ \partial \omega} , 
\label{dedt}
\\
\frac{d \Omega}{dt} & = & \frac{1}{n a^2 \sqrt{1-e^2} \sin I} 
\frac{\partial {\cal R}}{ \partial I} , 
\label{dOmegadt}
\\
\frac{d \omega}{dt} & = & \frac{1}{n a^2} \left( 
\frac{\sqrt{1-e^2}}{e} \frac{\partial {\cal R}}{ \partial e}
- \frac{\cot I}{\sqrt{1-e^2}} \frac{\partial {\cal R}}{ \partial I} \right),
\label{dwdt}
\\
\frac{d I}{dt} & = & \frac{ 1}{n a^2 \sqrt{1-e^2}} \left(
\cot I \frac{\partial {\cal R}}{ \partial \omega} -
\frac{1}{\sin I} \frac{\partial {\cal R}}{ \partial \Omega} \right),
\label{dIdt}
\end{eqnarray}

\noindent where ${\cal R} = - \left< \Phi_d \right>$ is the disturbing
function and $n=\sqrt{G(\Ms + M_p)/a^3}$ is the planet's mean motion.
From these equations and the expression~(\ref{potential3}) of
$\Phi_d$, a natural timescale emerges:

\be
T = n^{-1} \frac{\Ms + M_p}{ \Sigma_0 a^2} \frac{R_o}{a}.
\label{T}
\ee

\noindent Note that, when the evolution of the orbital elements is
periodic, in general $T$ is not the period of these oscillations (see
section~\ref{sec:solutions}).  However, equation~(\ref{T}) is
important as it gives the scaling of the evolutionary timescale with
the different parameters.  We see in particular that $T \propto
a^{-3/2} \Md^{-1}$.

Having expanded ${\cal R}$ to fourth order in $e$, we obtain $(\partial
{\cal R} / \partial e ) /e$ to second order in $e$.  We insert the
time--averaged expressions of $r_p^2$, $r_p^2 \cos 2 \theta_p$,
$z_p^2$ and $r_p z_p \sin \theta_p$ given by
equations~(\ref{rp2})--(\ref{rpzpsin}) into the
potential~(\ref{potential3}).  We then perform the derivatives of
${\cal R}= - \left< \Phi_d \right>$ with respect to $e$, $\Omega$,
$\omega$ and $I$ to obtain the dimensionless form of the Lagrangian
equations of motion~(\ref{dedt})--(\ref{dIdt}) to second order in $e$:

\begin{eqnarray}
  T \frac{de}{dt} = e  & & \left\{ \left( {\cal I}_2 - {\cal I}_4 \right) 
    \frac{5}{2} \sin 2 \omega \sin^2 I \right. \nonumber \\
  & & \left. + \frac{5 {\cal I}_3}{4} \left[ \left( 3+ \cos 2I \right) 
      \sin 2 \omega
      \cos 2 \Omega + 4 \cos I \cos 2 \omega \sin 2 \Omega \right] \right. 
  \nonumber \\
  & & \left.  -
    \frac{5 {\cal I}_5}{2} \sin I \left( \cos I \sin 2 \omega \cos \Omega
      + \cos 2 \omega \sin \Omega \right)   \right\} ,
\label{Tdedt}
\end{eqnarray}

\begin{eqnarray}
  T \frac{d \Omega}{dt} = & & \left( - {\cal I}_2+{\cal I}_4 \right)  
  \frac{\cos I}{2} \left( 2 + 4 e^2 -5 e^2 
    \cos 2 \omega \right) \nonumber \\ & + & {\cal I}_3 
  \left[ \frac{1}{2} \cos I 
    \left( 2 + 4 e^2 - 5 e^2 \cos 2 \omega \right) \cos 2 \Omega 
    + \frac{5}{2} e^2 \sin 2 \omega \sin 2 \Omega \right]
  \nonumber \\
  & + & \frac{{\cal I}_5}{4 \sin I} \left[ \left( 2 + 4 e^2 - 5 e^2 
\cos 2 \omega \right)
    \cos 2 I \cos \Omega  + 5 e^2 \cos I \sin 2 \omega \sin \Omega \right],
\label{TdOmegadt}
\end{eqnarray}

\begin{eqnarray}
  T \frac{d \omega}{dt} = & &
  \left( {\cal I}_2 - {\cal I}_4 \right) \frac{1}{8} \left[
    5 \left( 2 -3e^2 \right) \cos 2 \omega +  
    10 \left( 2 + e^2 \right)
    \cos 2 I \sin^2 \omega \right] \nonumber \\
  & + &  \frac{ {\cal I}_2}{8} \left( 22-e^2 \right)+ \frac{ {\cal I}_4}{8} 
  \left( 2-11e^2 \right)
  \nonumber \\
  & + & {\cal I}_3 \left\{ \frac{1}{8} \cos 2 \Omega \left[2 -11e^2
     +5 \left( 6 - e^2 \right)  \cos 2 \omega 
      - 10 \left( 2 + e^2 \right) \cos 2I \sin^2 \omega \right]  \right.
  \nonumber \\ 
  & & \left. \hspace{9.cm} -  5 
    \cos I \sin 2 \omega \sin 2 \Omega \right\} \nonumber \\
  & + & \frac{{\cal I}_5}{8 \sin I} \left\{ 
    \cos \Omega \cos I
    \left[ \left( -2 + e^2  \right) \left( -3 + 5 \cos 2 \omega \right) 
      -10 \left( 2+ e^2  \right) \cos 2I \sin^2 \omega \right] 
  \right.
  \nonumber \\
  &  & \left. \hspace{6.8cm} + 10 \left( 1 - e^2 - \cos 2I \right) 
    \sin 2 \omega \sin \Omega
  \right\}
\label{Tdwdt}
\end{eqnarray}

\begin{eqnarray}
  T \frac{d I}{dt} = & & \left( -{\cal I}_2 + {\cal I}_4 \right) \frac{5}{4}
e^2 \sin 2I  \sin 2 \omega \nonumber \\
& + & {\cal I}_3 \frac{ \sin I}{2} \left[ 
5 e^2 \cos I \cos 2 \Omega \sin 2 \omega 
+ \left( 2 + 4 e^2 + 5 e^2 \cos 2 \omega  \right) \sin 2 \Omega \right]
\nonumber \\
& + & \frac{{\cal I}_5 }{4} \left[ 5 e^2 \cos 2I \cos \Omega \sin 2 \omega
+ \cos I \left( 2 + 4e^2 + 5 e^2 \cos 2 \omega \right) \sin \Omega \right]
\label{TdIdt}
\end{eqnarray}

The perturbing function depends on the third Delaunay variable
$\Omega$.  Thus the Delaunay momentum, which is $\propto \sqrt{1-e^2}
\cos I$, is generally not constant.  This is in contrast to the
classical Kozai case (e.g., Innanen et al. 1997 and see
section~\ref{sec:Kozai} below).

\section{Solution of the secular Lagrangian equations: orbital evolution}
\label{sec:solutions}

We solve the set of coupled differential
equations~(\ref{Tdedt})--(\ref{TdIdt}) to obtain the evolution of the
orbital elements when the planet is subject to the secular
gravitational perturbation by the disc.  We first focus on the case of
a flat disc and then investigate the case of a finite warp.

\subsection{Case of a flat disc}
\label{sec:Kozai}

The flat disc case is obtained by taking $\tan \alpha=0$ in
equation~(\ref{Ho}).  Therefore, $m=0$ in
equations~(\ref{F1})--(\ref{F4}).  Since $F_1(0)=F_2(0)=F_3(0)= \pi
/2$ and $F_4(0)=0$, equations~(\ref{I2})--(\ref{I5}) give:

\be {\cal I}_2 = \frac{\pi}{2} \int_{\rho_i}^1 \frac{ \Sigma
  (\rho)}{\Sigma_0} \rho^{-2} d \rho = \frac{\pi}{2(q+1)} \left(
  \rho_i^{-q-1} - 1 \right),
\label{I2F}
\ee

\noindent ${\cal I}_3 = {\cal I}_5 = 0$ and ${\cal I}_4 = -2 {\cal
  I}_2$.  In this limit, the potential given by
equation~(\ref{potential3}) reduces to the following expression:

\be \left< \Phi_d \right> = - \frac{G \Sigma_0}{R_o} \left[ {\cal I}_1
  R_o^2 + 2 {\cal I}_2 \left( \frac{1}{2} \left< r_p^2 \right> -
    \left< z_p^2 \right> \right) \right], \ee

\noindent which is of the same form as in the classical Kozai case
(Kozai 1962, Lidov 1962), in which an inner body is perturbed by a
distant companion on an inclined orbit.  When only the secular terms
are considered, the Kozai case is equivalent to averaging the mass of
the outer companion over its orbit.  As pointed out by Terquem \&
Ajmia (2010), this is analogous to having the inner body perturbed by
a disc, as long as most of the disc's mass is beyond the orbit of the
body.  We therefore expect that the perturbing potential will have the
same form in both cases.  Note that here $\left< \Phi_d \right>$ does
not depend on the third Delaunay variable $\Omega$, and therefore
$\sqrt{1-e^2} \cos I$ is constant.

With the above values of the ${\cal I}_i$, the Lagragian equations of
motion~(\ref{Tdedt})--(\ref{TdIdt}) reduce to:

\begin{eqnarray}
T \frac{de}{dt} & = &   \frac{15}{2} {\cal I}_2 e \sin 2 \omega \sin^2 I , 
\label{TdedtF}
\\
T \frac{d \Omega}{dt} & = & -  \frac{3}{2} {\cal I}_2 \cos I \left(
2 + 4 e^2 - 5 e^2 \cos 2 \omega \right),
\label{TdOmdtF} 
\\
T \frac{d \omega}{dt} & = & \frac{3}{8} {\cal I}_2 
\left[ 6 + 7 e^2 + 5 \left( 2 - 3 e^2 \right) \cos 2 \omega + 
   10 \left( 2 + e^2 \right) \cos 2I \sin^2 \omega  \right],
\label{TdwdtF} 
\\
T \frac{dI}{dt} & = &  -  \frac{15}{4} {\cal I}_2  e^2 \sin 2 \omega \sin 2I .
\label{TdIdtF}
\end{eqnarray}

\noindent To second order in $e$, these
equations are the same as equations~(5) from Innanen et al. (1997),
which were derived in the classical Kozai case, provided that we correct an
error in their expression for $d \omega / d \tau$, in which the factor
$\sqrt{1-e^2}$ should be in the denominator, not in the numerator.

\begin{figure}
\begin{center}
\includegraphics[scale=0.7]{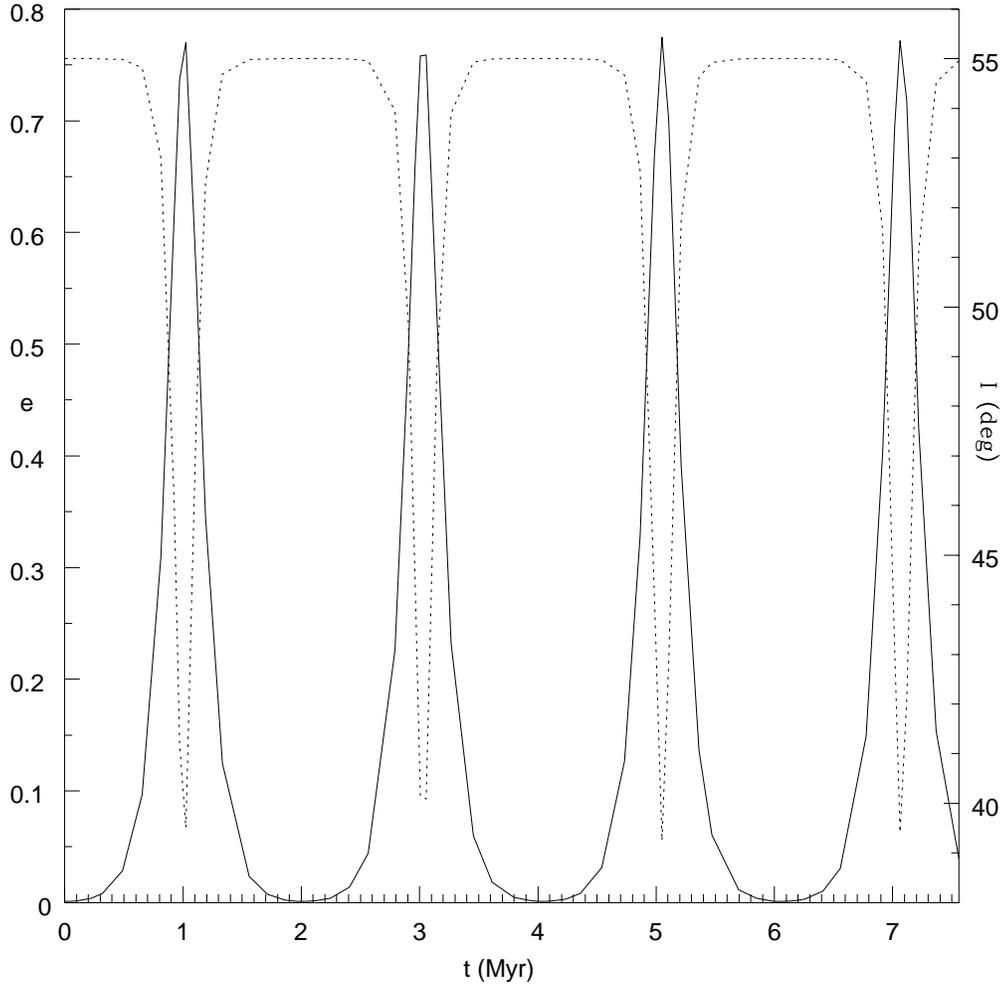}
\end{center}
\caption{Eccentricity $e$ (solid line) and inclination angle $I$ (in
  degrees, dotted line) versus time $t$ in Myr for $\alpha=0$ (flat
  disc), $a=10$~au, $\Mp =10^{-3}$~M$_{\odot}$, $\Md
  =10^{-2}$~M$_{\odot}$, $R_i = 60$~au, $R_o = 80$~au, $e_0=10^{-3}$
  and $I_0=55^{\circ}$.  This illustrates the classical Kozai case, in
  which $e$ and $I$ undergo antiphase oscillations.  }
\label{fig1}
\end{figure}

Figure~\ref{fig1} shows $e$ and $I$ versus time, obtained by solving
equations~(\ref{TdedtF})--(\ref{TdIdtF}) with the following initial
values: $e_0=10^{-3}$, $\Omega_0=\omega_0=0$ and $I_0=55^{\circ}$.  (In
this section, we only aim at showing that the equations we use reduce
to the classical Kozai case in the appropriate limit, and thefore we
do not address the issue of how a planet could start on an inclined
orbit.)  We have fixed $q=1/2$ in the expression~(\ref{sigma}) of
$\Sigma$, $R_i=60$~au, $R_o=80$~au, $\Md =10^{-2}$~M$_{\odot}$,
$a=10$~au and $\Mp =10^{-3}$~M$_{\odot}$.  For these parameters,
equation~(\ref{I2F}) gives ${\cal I}_2=0.565$ and the 
timescale~(\ref{T}) is $T=3.8 \times 10^5$~years for
$\Ms=1$~M$_\odot$.  As expected, since $I_0$ is larger than the
critical angle $I_c= \arccos \sqrt{3/5} =39^{\circ}.23$, $e$ and $I$
undergo antiphase oscillations (Kozai 1962, Innanen et al. 1997).
Note that $T$ is not the period of these oscillations.  The
inclination angle varies between $I_c$ and $I_0$.  In theory, as
$\sqrt{1-e^2} \cos I$ is constant, the maximum eccentricity should be
$\sqrt{1-5 \cos^2 I_0/3} = 0.67$.  We see in figure~\ref{fig1} that
the maximum eccentricity is actually a bit larger.  This is because
when $e$ reaches the largest values, the analysis above, which is
valid only to second order in $e$, breaks down and $\sqrt{1-e^2} \cos
I$ is not constant (it varies by about 14\% is this particular case).

As noted, $p=1$ in equation~(\ref{zwarp}) corresponds to a disc that
is rigidly tilted with respect to the $(x,y)$--plane.  In this case,
the term proportional to $\sin \theta$ in the denominator of the
integral over $\theta$ in the potential~(\ref{potential}) can be
eliminated by the appropriate choice of coordinates, and this reduces
to the Kozai case.  Here however, we have defined the orbital elements
with respect to the $(x,y)$--plane and {\em not} to the plane in which
the disc lies when $p=1$, so we cannot simply recover Kozai's results
when $p=1$.

\subsection{Case of a warped disc}

To investigate the warped disc case ($\alpha \ne 0$), we solve
equations~(\ref{Tdedt})--(\ref{TdIdt}).  Here again we fix $q=1/2$ in
the expression~(\ref{sigma}) of $\Sigma$ and we adopt $p=2$ or 3 in
equation~(\ref{zwarp}).  In principle, the analysis in this section is
valid only for $a \ll R_i$.  However, as we will see below, some of
the results in this section do not depend on $a$, and as it happens, any
value of $a < R_i$ could be considered.

When the disc reduces to a narrow ring ($R_o$ only slightly larger
than $R_i$), we expect the orbital angular momentum vector ${\bf
  L}_{\rm orb}$ to precess around the total angular momentum vector.
Since the ring's angular momentum is large compared to that of the
orbit, this means that ${\bf L}_{\rm orb}$ precesses around the ring's
angular momentum vector ${\bf L}_{\rm ring}$.  As the ring is inclined
by an angle $\alpha$ with respect to the $(x,y)$--plane, the
inclination $I$ of the orbit varies periodically between its initial
value $I_0$ and $I_{\rm max}=2 \alpha-I_0$ during the precession,
independently of $a$ and $\Mp$.

\begin{figure}
\begin{center}
\includegraphics[scale=0.7]{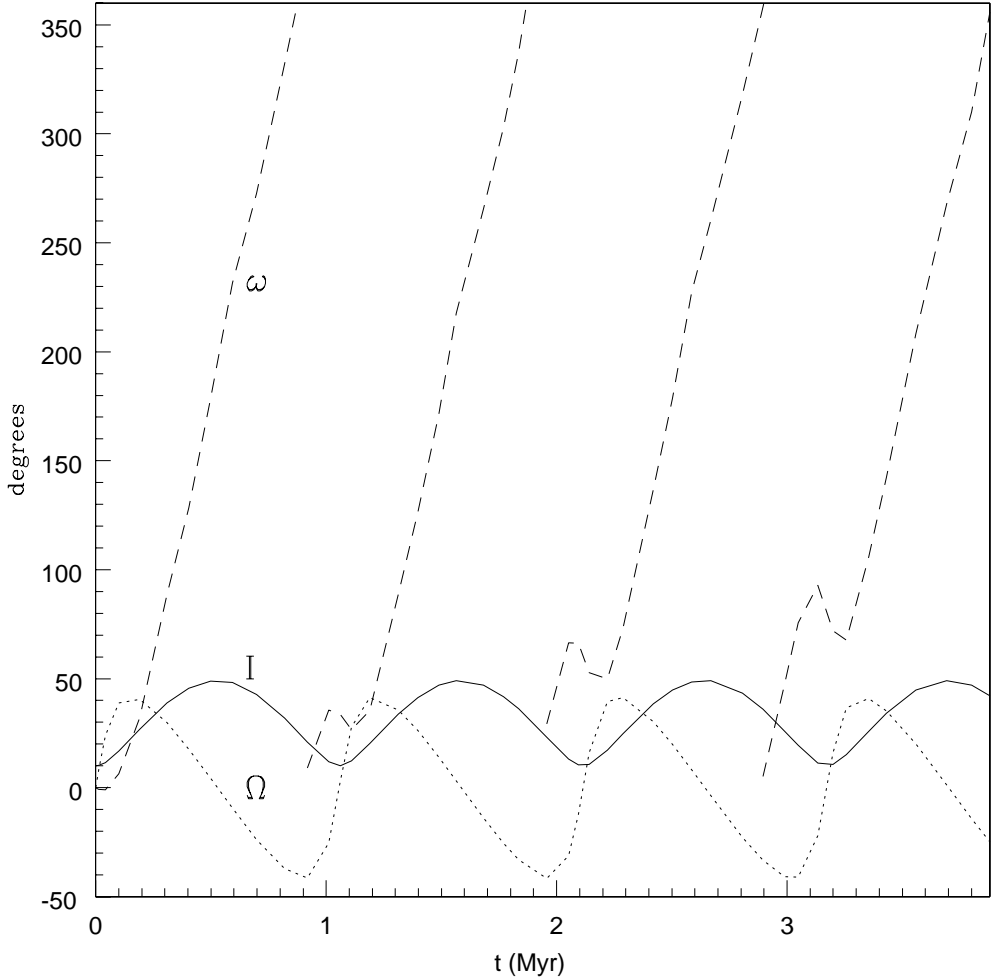}
\end{center}
\caption{Inclination angle $I$ (solid line), longitude of ascending
  node $\Omega$ (dotted line) and argument of pericentre $\omega$
  (dashed line) versus time $t$ in Myr for $\alpha=30^{\circ}$,
  $p=3$, $a=10$~au, $\Mp =10^{-3}$~M$_{\odot}$, $\Md
  =10^{-2}$~M$_{\odot}$, $R_i = 60$~au, $R_o = 61$~au, $e_0=10^{-3}$
  and $I_0=10^{\circ}$.  The orbital angular momentum vector precesses
  around the ring's angular momentum vector, which causes $I$ to
  oscillate bewteen 10 and $50^{\circ}$ and $\omega$ to circulate.}
\label{fig2}
\end{figure}

This is illustrated in figure~\ref{fig2}, which shows $I$, $\omega$
and $\Omega$ versus time for the initial values $e_0=10^{-3}$,
$I_0=10^{\circ}$ and $\Omega_0=\omega_0=0$, and for $a=10$~au, $\Mp
=10^{-3}$~M$_{\odot}$, $\alpha=30^{\circ}$, $p=3$, $\Md
=10^{-2}$~M$_{\odot}$, $R_i = 60$~au and $R_o = 61$~au.  We have
chosen a finite value of $I_0$ because, as equations~(\ref{TdOmegadt})
and~(\ref{Tdwdt}) show, $d\Omega/dt$ and $d\omega/dt$ become infinite
when $I \rightarrow 0$.  However, we have checked that exactly the
same variations of $\Omega$ and $I$ are obtained for $I_0=1^{\circ}$,
for example.  In this latter case, only the variations of $\omega$ are a bit
more irregular around the times when $I$ is minimum.  As we had
anticipated, $I_{\rm max}=50^{\circ}= 2 \alpha-I_0$.  Also, as
expected for a precessing motion, $\omega$ circulates uniformly.  As
can be seen from equation~(\ref{Tdedt}), this implies that changes in
the eccentricity due to the torque exerted by the disc average to zero
over a precession period, so that no secular change of the
eccentricity occurs.

Note that exactly the same curves are obtained for different values of
$a$, $\Mp$ or $\Md$.  Only the precession period $T_{\rm prec} \propto
a^{-3/2} \Md^{-1}$ changes when these parameters are varied.

In the case displayed in figure~\ref{fig2}, the ring is narrow and
essentially all of its mass is in a plane inclined by $\alpha$ with
respect to the $(x,y)$--plane.  If we now increase the ratio
$R_o/R_i$, while keeping $\Md$ fixed, more of the disc's mass will be
at smaller inclination with respect to the $(x,y)$--plane, and
therefore the angle between ${\bf L}_{\rm ring}$ and ${\bf L}_{\rm
  orb}$ will be smaller.  The maximum value reached by $I$ during the
precession will accordingly be reduced.  For the same parameters as
above but with $R_o=80$ or 100~au, we obtain $I_{\rm max}=36^{\circ}$ or
$27^{\circ}$, respectively.  If we keep $R_0=61$~au but take
$R_i=50$~au, $I_{\rm max}$ is reduced to $40^{\circ}$.  These values
of $I_{\rm max}$ would be higher for $p=2$, as in that case the inner
parts of the disc would have a higher inclination (see table below).

From equations~(\ref{zwarp}) and~(\ref{Ho}), we see that the disc can
be modelled as a collection of rings of width $dr$ located at a
distance $r$ from the central star and inclined with respect to the
$(x,y)$--plane by an angle:

\be \beta(r)=\arctan \left[ \left( \frac{r}{R_o} \right)^{p-1} \tan
  \alpha \right]. \ee

\noindent The planet will be more affected by rings which are more
massive and closer.  We therefore define an average inclination angle
$\overline{\beta}$ over the whole disc by weighting $\beta(r)$ by the
mass of the ring and a function $f(r)$ that decreases with $r$ and
such that $f(R_i)=1$:

\be
\overline{\beta} = \frac{1}{\Md} \int_{R_i}^{R_o} 2 \pi r dr \Sigma(r) f(r)
\beta(r).
\ee

\noindent We would expect the orbital angular momentum vector to
precess around the vector inclined by this angle with respect to the
vertical axis, and therefore $I_{\rm max}= 2 \overline{\beta} - I_0$.
We have found that this is the case to within a good approximation if
we take $f(r)=\sqrt{R_i/r}$, which we will adopt thereafter.  For the
parameters used above, with $R_i=60$~au and $R_o=80$ or 100~au, we obtain
$\overline{\beta}=22^{\circ}$ or $18^{\circ}$, respectively, and
therefore $2 \overline{\beta} - I_0= 34^{\circ}$ or $26^{\circ}$,
reasonably close to the values of $I_{\rm max}$ which were found to be
$36^{\circ}$ and $27^{\circ}$.

Table~\ref{table} gives $I_{\rm max}$ and $\overline{\beta}$ for
different values of $\alpha$, $R_i$ and $R_o$ and for $p=2$ or 3.  In
this table, we have considered parameters for which the interaction
between the disc and the planet leads to a precession of the orbit.
As noted above, $I_{\rm max}$ does not depend on $a$, and therefore
the results listed here are valid for any $a < R_i$, not just $a \ll
R_i$.  This has been checked by doing numerical simulations using the
exact disc potential, i.e. without assuming that $a$ is very small
compared to $R_i$ (see section~\ref{sec:simulations}).

We use equation~(\ref{T}) to write the precession period $T_{\rm
  prec}$ as a function of $a$ and $\Md$ and define $\overline{T}_{\rm
  prec}$ such that:

\be
 T_{\rm prec} = \overline{T}_{\rm prec} \left( 
\frac{ 1 \; {\rm au}}{a} \right)^{3/2} 
\frac{ 10 \; {\rm M}_{\rm J}}{\Md} .
\ee

\noindent 
$\overline{T}_{\rm prec}$ is given in table~\ref{table} for the cases
considered and is equal to the precession period for $a=1$~au and
$\Md=10$~$\MJ$.  Note that this precession period is divided by about
10 if we take $a=5$~au.  Neither $I_{\rm max}$ nor $T_{\rm prec}$
depend on $\Mp$.  For the initial inclination angle $I_0=1^{\circ}$
used here, we expect $I_{\rm max} = 2 \overline{\beta}$.  We note that
this is approximately the case, except when $R_o/R_i$ is large and
$p=3$, which corresponds to a disc with very low inclinations in the
inner parts.

\begin{table}
  \caption{Maximum inclination and precession period  
in the regime where the orbit precesses for $p=3$ and $p=2$.}
\label{table}
\begin{tabular}{ccccccccccc}
  \cline{5-7}
\cline{9-11} 
  & & & & \multicolumn{3}{c}{p=3} & & \multicolumn{3}{c}{p=2} \\
  \hline
  $\alpha$ &  $R_i$  & $R_o$ & &  $\overline{\beta}$ & $I_{\rm max}$ & 
  $\overline{T}_{\rm prec}$ & &  $\overline{\beta}$ & $I_{\rm max}$ & 
  $\overline{T}_{\rm prec}$ \\
  (degrees) & (au) & (au)  & & (degrees) & (degrees) & (Myr) & &
  (degrees) & (degrees) & (Myr) \\
  \hline
  30  & 60  & 61  & & 29.5 & 59 & 36  & & 29.5 & 59 & 36  \\
  --- & --- & 80  & &  22  & 45 & 50  & & 25   & 52 & 52  \\
  --- & --- & 100 & &  18  & 36 & 68  & & 21.5 & 46 & 70  \\
  --- & --- & 150 & & 12.5 & 23 & 120 & & 16.5 & 36 & 124  \\
  --- & --- & 200 & &  10  & 16 & 180 & & 14   & 30 & 187  \\
  30  & 15  & 100 & & 6.5  & 6  & 7.8 & & 9.5  & 18 & 8.0  \\
  50  & --- & --- & & 12   & 12 & 8.1 & & 17.5 & 35 & 8.6  \\
  70  & --- & --- & & 21   & 24 & 8.8 & & 28.5 & 68 & 11  \\
  \hline
\end{tabular} \\
\medskip
The initial conditions are $I_0=1^{\circ}$, $e_0=10^{-3}$ and 
$\Omega_0=\omega_0=0$.  $I_{\rm max}$ is independant of $\Md$, $\Mp$ and 
$a$ as long as $a<R_i$.  The precession period is $T_{\rm prec}= 
\overline{T}_{\rm prec} \times (1 \; {\rm au} / a)^{3/2} 
(10 \; {\rm M}_{\rm J} / \Md)$.
\end{table}

\begin{figure}
\begin{center}
\includegraphics[scale=0.7]{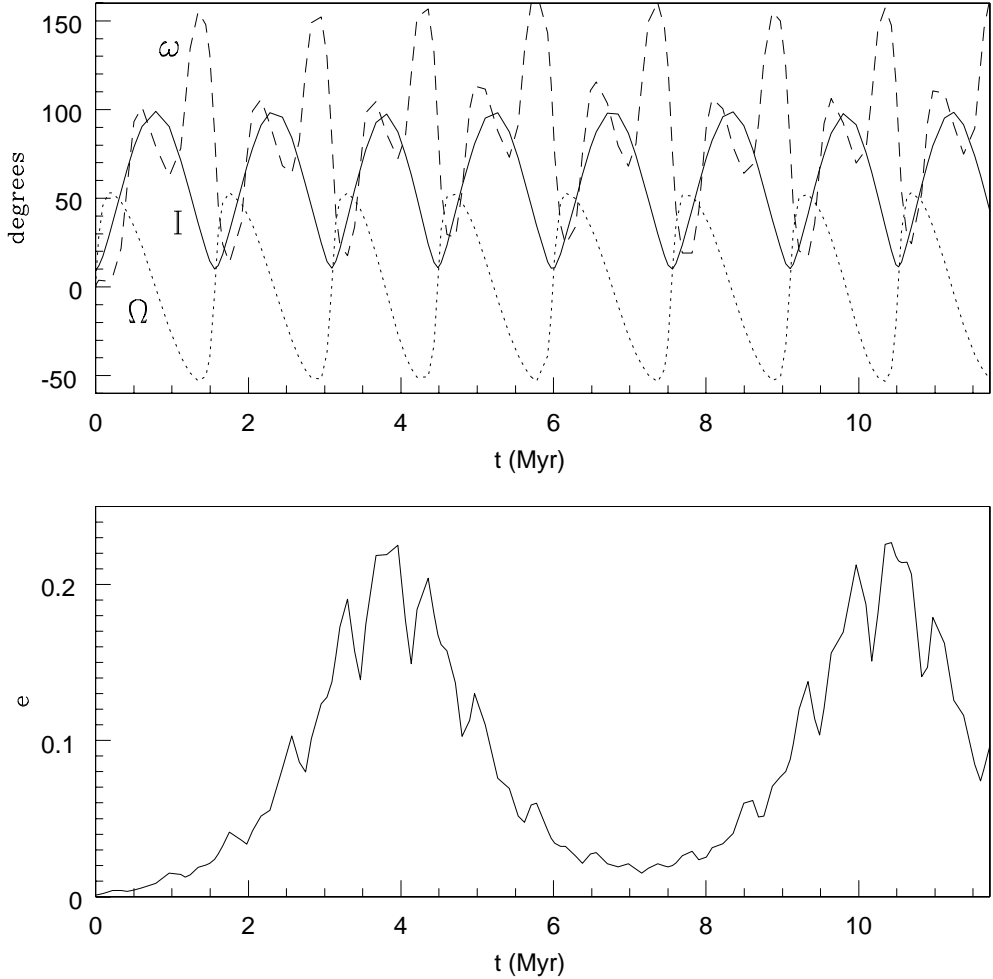}
\end{center}
\caption{Inclination angle $I$ (solid line, upper plot), longitude of
  ascending node $\Omega$ (dotted line, upper plot), argument of
  pericentre $\omega$ (dashed line, upper plot) and eccentricity $e$
  (lower plot) versus time $t$ in Myr for the same parameters as
  figure~\ref{fig2} except for $\alpha=55^{\circ}$.  When $I$
  approaches $90^{\circ}$, $\omega$ no longer circulates and $e$
  grows.  }
\label{fig3}
\end{figure}

When $I_{\rm max}$ gets larger than $90^{\circ}$, the orbit becomes
retrograde (more precisely, it would become retrograde relative to the
stellar spin if stellar rotation were taken into account).  In the
case of the narrow ring and for $I_0=10^{\circ}$, this happens for
$\alpha > 50^{\circ}$.  For $\alpha=49^{\circ}$, $I_{\rm
  max}=88^{\circ}$ and $e$ stays small (smaller than $3 \times
10^{-3}$).  In that case, $\omega$ still circulates, although not as
smoothly as for smaller values of $\alpha$.  For $\alpha=55^{\circ}$,
$I_{\rm max}=100^{\circ}$ and $e$ grows to 0.23.  This is illustrated
in figure~\ref{fig3}, which shows $I$, $\omega$, $\Omega$ and $e$ for
the same parameters as figure~\ref{fig2} except for
$\alpha=55^{\circ}$.  At the beginning of the evolution, $\omega$
increases .  But as $I$ approaches $90^{\circ}$, $\omega$ no longer
circulates, so that changes in $e$ no longer average to zero over time
and a net increase of the eccentricity is observed.

As $\alpha$ is increased, $e$ becomes too large for the second order
analysis of this section to apply.

\section{Numerical simulations}
\label{sec:simulations}

The analysis presented in the above section is valid only for small
values of $e$.  In principle, it is also valid only for $a \ll R_i$.
However, as pointed out in the previous section, when the planetary
orbit precesses, the variations of $I$ do not depend on $a$ and
therefore any value of $a$ may be considered in the analysis.  To
study the evolution of the planet's orbit in a more general case, we
use the $N$--body code described in Papaloizou~\& Terquem~(2001) in
which we have added the gravitational force exerted by the disc onto
the planet.

The equation of motion for the planet is:
\begin{equation} 
{d^2 {\bf r} \over dt^2} = -{GM_\star{\bf r} \over |{\bf r}|^3} -
\mbox{\boldmath $\nabla$} \Phi_d - {GM_p{\bf r} \over |{\bf r}|^3} + {\bf
\Gamma}_{t,r} \; , \label{emot}
\end{equation} 

\noindent where ${\bf r}$ is the position vector of the planet and
$\Phi_d$ is the gravitational potential of the disc given by
equation~(\ref{potential}).  Note that here we use the exact disc
potential, so that the results of this section are valid in the
general case $a<R_i$, not just $a \ll R_i$.  The third term on the
right--hand side is the acceleration of the coordinate system based on
the central star.  This neglects the net force of the disc on the
central star, which would of course require a calculation of the disc
response to the planet's perturbation.  Tides raised by the star in
the planet and relativistic effects are included through ${\bf
  \Gamma}_{t,r}$, but they are unimportant here as the planet does not
approach the star closely.  Equation~(\ref{emot}) is integrated using
the Bulirsch--Stoer method and the integrals involved in
$\mbox{\boldmath $\nabla$} \Phi_d$ are calculated with the Romberg
method (Press et al.  1993).

The planet is set on a circular orbit in the $(x,y)$--plane at a
distance $a$ from the star.  As here there is no frictional force
acting on the planet, the orbital energy is conserved and the
semimajor axis $a$ is constant throughout the evolution of the system.
In this paper, we suppose that the planet's orbit is inside the disc
inner cavity ($a<R_i$).  If the orbit were still embedded in the disc,
the frictional force exerted by the disc on the planet would damp the
planet's inclination and eccentricity, and this would have to be taken
into account.

In the simulations reported here, we have taken $q=1/2$ in the
expression~(\ref{sigma}) of $\Sigma$, $p=3$ in equation~(\ref{zwarp})
and $\Ms=1$~M$_{\odot}$, $\Mp =10^{-3}$~M$_{\odot}$ and $\Md
=10^{-2}$~M$_{\odot}$.

\begin{figure}
\begin{center}
\includegraphics[scale=0.7]{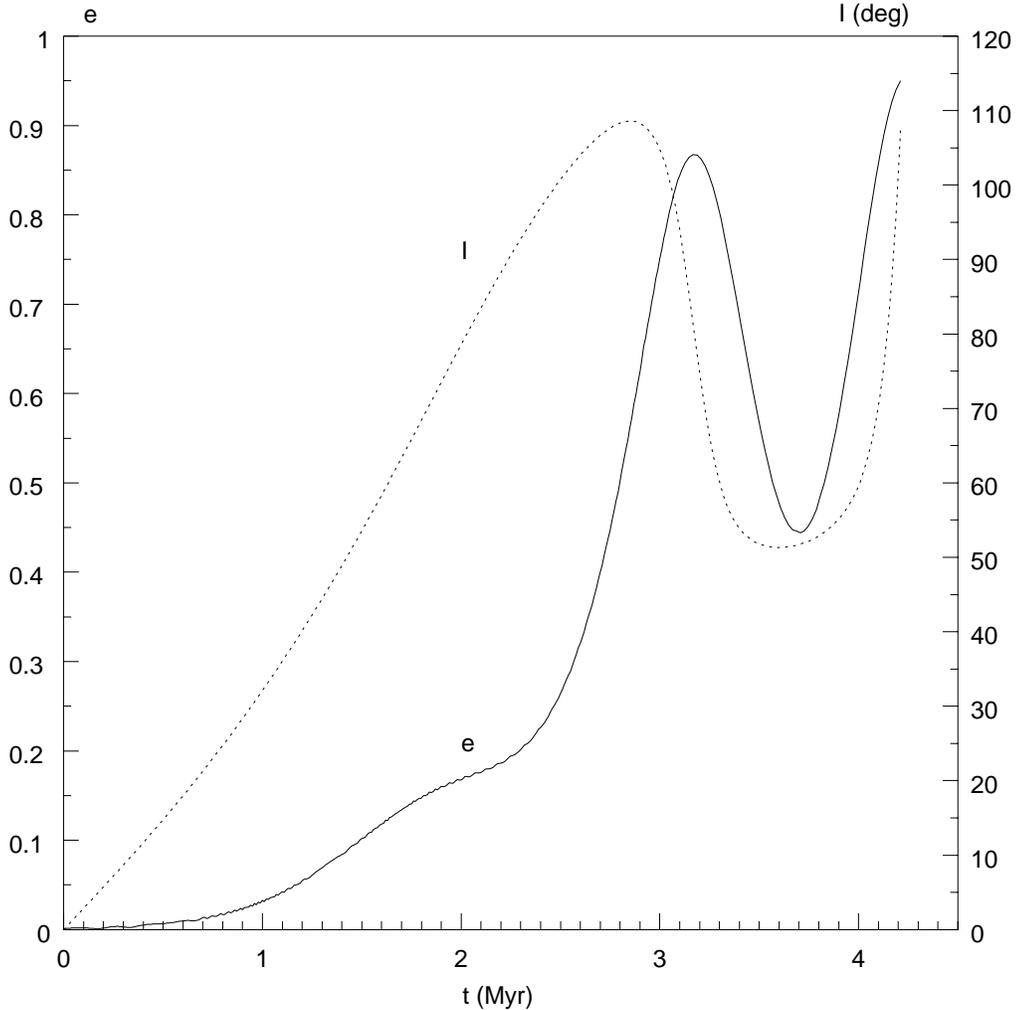}
\end{center}
\caption{Eccentricity $e$ (solid line) and inclination angle $I$ (in
  degrees, dotted line) versus time $t$ in Myr for
  $\alpha=70^{\circ}$, $a=10$~au, $\Mp =10^{-3}$~M$_{\odot}$, $\Md
  =10^{-2}$~M$_{\odot}$, $R_i = 60$~au and $R_o = 80$~au. The orbit
  becomes retrograde and the eccentricity is increased to very high
  values.  }
\label{fig4}
\end{figure}

Figure~(\ref{fig4}) shows the evolution of $e$ and $I$ for
$\alpha=70^{\circ}$, $a=10$~au, $R_i = 60$~au and $R_o = 80$~au.  Note
that exactly the same evolution is obtained for different values of
$\Mp$.  The orbit becomes retrograde and the eccentricity is pumped up
to very high values.  We have stopped the calculation when, due to the
very high eccentricity, the code failed to conserve the energy to
within a specified accuracy .  As $e$ approaches unity, we expect
tidal interaction with the central star to become important and
dissipate orbital energy so that $e$ decreases.  Ultimately, if the
disc does not disappear, the orbit would become circularized on a
short orbit.  We have verified that we obtain a similar evolution by
solving equations~(\ref{Tdedt})--(\ref{TdIdt}) in the early stages
when $e \ll 1$.

We observe exactly the same variations of $I$ and $e$ if we take
$a=20$~au instead of $a=10$~au.  The evolution is about three
times faster for this wider orbit, which is consistent with an
evolutionary timescale propotional to $a^{-3/2}$.  Similar
behaviour is observed for $p=2$ (instead of $p=3$ used here), also on
a shorter timescale.

Similar evolution is observed for other values of $\alpha$.  The case
displayed in figure~\ref{fig4} is typical, illustrating
what happens when the inclination of the orbit passes through
$90^{\circ}$: the eccentricity is pumped up to high values and $I$
varies over a broad range of values, including retrograde
inclinations.

\section{Discussion}
\label{sec:discussion}

In this paper, we have studied the secular interaction between a
planet and a warped disc.  We have considered the stage of evolution
after the planet has cleared--out a gap.  Therefore, there is no
friction between the disc and the planet, and the planetary orbit
evolves only under the effect of the gravitational potential from the
disc.  In general, we find that the interaction between the planet and
the disc leads to the precession of the orbit.  The orbital plane
therefore becomes inclined relative to the plane which is tangent to
the disc's inner parts, which we will hereafter call the ``reference
plane''.  If we define some average inclination of the disc with
respect to the reference plane, then the inclination of the orbital
plane is twice this averaged angle.  The precession of the orbit does
not modify its eccentricity: if the planet started on a circular
orbit, the eccentricity stays small.

There is an instability when the inclination of the orbital plane
relative to the reference plane approaches 90$^{\circ}$.  The
eccentricity of the planet is then built up and both the inclination
and the eccentricity develop very large amplitude variations.  That
happens when the disc is very severely warped, or if there is a
significant amount of mass in a ring inclined by a least 45$^{\circ}$
relative to the initial orbital plane.

Note that the inclination of the orbital plane is measured relative to
the disc's inner region.  The disc as a whole will in general precess
around the central star's equatorial plane, and this will have to be
taken into account to compute the inclination of the orbital plane
relative to this equatorial plane.

The process studied in this paper shows that the orbit of a planet
that has formed in a warped disc could develop a significant
inclination when there is no frictional damping. 

Although the inclination of the orbital plane does not depend on the
planet's semimajor axis $a$, the timescale for generating a
significant inclination does increase as $a$ decreases.  For a disc of
$\sim 10$~$\MJ$, we typically find that for this timescale to be at
most of a few Myr, $a$ has to be larger than 5~au.  We therefore
expect only planets beyond the snow line to become significantly
misaligned through this process.  If the warp is such that the orbital
plane precesses, the eccentricity is not built up and the planet would
be left on its original semimajor axis as the disc dissipates.
However, if large eccentricities are generated, the pericentre would
become small enough for tidal interaction with the star to circularize
the orbit.  The planet would then end up on a short orbit, as are the
misaligned planets observed so far.  

The process described in this paper therefore leads to two distinct
populations of misaligned planets: one with objects beyond the snow
line and at most moderate eccentricities, and another with objects on
short and circular orbits.  Of course, the presence of other planets in
the system would modify this picture as planet--planet interactions
would then change the orbital elements.  So far, only planets on short
orbits have been detected on inclined orbits.  Measurements of the
projected inclination angle for more distant objects would help to put
constraints on the process discussed here.

\section*{Acknowledgements}

I would like to thank J. Silk for suggesting that the interaction
between a planet and a warped disc may drive the inclination of the
planetary orbit.  It is a pleasure to thank S. Balbus for a critical
reading of the manuscript and useful comments.

%
%


%

\appendix

\section{Expression  of the gravitational potential due to the disc
 in terms of  elliptic  integrals}
\label{appendixA}

Using the definition of ${\cal I}(r,u,v)$ given by
equation~(\ref{Idefinition}), we can calculate the different terms
that enter its expansion~(\ref{Iexpansion}):

\be
{\cal I}(r,0,0)  =  4 F_1(m), 
\ee
\be
\frac{\partial^2 {\cal I}}{\partial u^2} \left( r,0,0 \right) 
 =   - 4 F_2(m) +  6 F_3(m) +  6 \cos 2 \theta_p F_4(m) , 
\ee
\be
\frac{\partial^2 {\cal I}}{\partial v^2} \left( r,0,0 \right) 
 =  - 4 F_2(m) + 6 m^2 \left[  F_3(m) - F_4(m) \right] , 
\ee
\be
\frac{\partial^2 {\cal I}}{\partial u v} \left( r,0,0 \right) 
 =  6 m \sin \theta_p \left[  F_3(m) - F_4(m) \right] ,
\ee

\noindent where we have defined, for $i=1$, 2 or 3:

\be 
F_i(m) = \int_0^{\pi/2} \frac{ d
    \theta}{\left( 1+m^2 \sin^2 \theta \right)^{(2i-1)/2}},
\ee

\noindent and:

\be
F_4(m) = \int_0^{\pi/2} \frac{ \cos 2 \theta d
    \theta}{\left( 1+m^2 \sin^2 \theta \right)^{5/2}}.
\label{F40}
\ee

\noindent We recall that $m=\left( r / R_o \right)^{p-1} \tan \alpha$.
The $F_i$ can be expressed in terms of the complete elliptic integrals
of the first and second kind, $K$ and $E$ respectively:

\be
F_1(m) = \frac{1}{\sqrt{1+m^2}} K \left( \frac{m^2}{1+m^2} \right) ,
\label{F1}
\ee

\be
F_2(m) = \frac{1}{\sqrt{1+m^2}} E \left( \frac{m^2}{1+m^2} \right) ,
\label{F2}
\ee

\be
F_3(m)= \frac{1}{3 \left( 1+m^2 \right)^{3/2}} 
\left[ 2 \left( 2 + m^2 \right) E \left( \frac{m^2}{1+m^2} \right)
- K \left( \frac{m^2}{1+m^2} \right) \right] ,
\label{F3}
\ee

\be
F_4(m)= \frac{1}{3 m^2 \left( 1+m^2 \right)^{3/2}} 
\left[ 2 \left( 1 + m^2 + m^4 \right) E \left( \frac{m^2}{1+m^2} \right)
- \left( 2 + m^2 \right) K \left( \frac{m^2}{1+m^2} \right) \right] ,
\label{F4}
\ee

\noindent where $K$ and $E$ are defined as:

\begin{eqnarray}
K(u) & = & \int_0^{\pi/2} \frac{d \theta}{\sqrt{1-u \sin^2 \theta}} , \\
E(u) & = & \int_0^{\pi/2} \sqrt{1-u \sin^2 \theta} d \theta,
\end{eqnarray}

\noindent with $u<1$.  Note that $F_4(0)$ should be calculated using
equation~(\ref{F40}) instead of equation~(\ref{F4}).

By inserting the expansion of ${\cal I}$ given by
equation~(\ref{Iexpansion}) into the expression of the potential
$\Phi_d$ given by equation~(\ref{potential2}), we then get the
potential under the form given by equation~(\ref{potential3}) with:

\begin{eqnarray}
{\cal I}_1 & = & \int_{\rho_i}^{1} 4 \overline{\Sigma}( \rho )
F_1(m) d \rho , 
\label{I1}
\\
{\cal I}_2 & = & \int_{\rho_i}^{1} \overline{\Sigma}( \rho )
\rho^{-2} \left[ -2 F_2(m) + 3 F_3(m) \right] d \rho , 
\label{I2}
\\
{\cal I}_3 & = & \int_{\rho_i}^{1} 3 \overline{\Sigma}( \rho )
\rho^{-2} F_4(m) d \rho , 
\label{I3}
\\
{\cal I}_4 & = & \int_{\rho_i}^{1}  \overline{\Sigma}( \rho )
\rho^{-2}
\left\{ -2  F_2(m) + 3 m^2  \left[
F_3(m) - F_4(m) \right] \right\} d \rho, 
\label{I4}
\\
{\cal I}_5 & = & \int_{\rho_i}^{1}  6 m \rho^{-2} 
\overline{\Sigma}( \rho ) \left[ F_3(m) - F_4(m) \right] d \rho
\label{I5}
\end{eqnarray}

\noindent with $\rho=r/R_o$, $\rho_i=R_i/R_o$ and $\overline{\Sigma}(
\rho ) = \Sigma( \rho ) / \Sigma_0$.  Note that $m= \rho^{p-1} \tan
\alpha$.

\section{Series expansion of $R$, $\sin f $ and $\cos f$ in terms of
  the mean anomaly to fourth order in the eccentricity }
\label{appendixB}

The series expansions of $R$, $\sin f$ and $\cos f$ in terms of $M$ to
fourth order in $e$ are given, e.g., in Murray \& Dermott (1999), and
we recall them below:

\begin{eqnarray}
\frac{R}{a} = 1 - e \cos M + \frac{e^2}{2} \left( 1 - \cos 2M \right)
& + & \frac{3 e^3}{8} \left( \cos M - \cos 3M \right) \nonumber \\
& + &  \frac{e^4}{3} \left( \cos 2M - \cos 4M \right) + {\cal O}(e^5),
\end{eqnarray}

\begin{eqnarray}
  \sin f = \sin M & + & e \sin 2M  +  \frac{e^2}{8} \left( 9 \sin 3M - 
7 \sin M \right)
  +  \frac{e^3}{3} \left( 4 \sin 4M - \frac{7}{2} \sin 2M \right) 
\nonumber \\
  & + &  \frac{e^4}{64} \left( \frac{17}{3} \sin M - \frac{207}{2} \sin 3M 
+\frac{625}{6} \sin 5M \right) + {\cal O}(e^5),
\end{eqnarray}

\begin{eqnarray}
  \cos f = \cos M & + & e \left( \cos 2M -1 \right)  + 
\frac{9 e^2}{8} \left( \cos 3M - \cos M \right)
  +  \frac{4e^3}{3} \left( \cos 4M - \cos 2M \right) 
\nonumber \\
  & + &  \frac{25 e^4}{64} \left( \frac{1}{3} \cos M - \frac{9}{2} \cos 3M 
+\frac{25}{6} \cos 5M \right) + {\cal O}(e^5).
\end{eqnarray}

\label{lastpage}
\end{document}